\documentclass[reprint,nofootinbib,amsmath,amssymb,aps]{revtex4-1}
\newcommand{\xx}{\mbox{\boldmath $x$}}
\usepackage{natbib}
\usepackage{color}
\begin{document}
\title{$\delta N$ formalism}
\author{Naonori S. Sugiyama}
\email{sugiyama@astr.tohoku.ac.jp}
\affiliation{Astronomical Institute, Graduate School of Science, Tohoku
University, Sendai 980-8578, Japan}  
\author{Eiichiro Komatsu}
\affiliation{Texas Cosmology Center and the Department of Astronomy,
The University of Texas at Austin, 1 University Station, C1400, Austin,
Texas 78712, USA}
\affiliation{Kavli Institute for the Physics and Mathematics of the
Universe, Todai Institutes for Advanced Study, The University of Tokyo,
Kashiwa, Japan 277-8583 (Kavli IPMU, WPI)}
\affiliation{Max Planck Institut f\"ur Astrophysik, Karl-Schwarzschild-Str. 1, 85741 Garching, Germany}
\author{Toshifumi Futamase}
\affiliation{ Astronomical Institute, Graduate School of Science, Tohoku
University, Sendai 980-8578, Japan}
\begin{abstract}
Precise understanding of nonlinear evolution of cosmological
 perturbations during inflation is necessary for the correct
 interpretation of measurements of non-Gaussian correlations in the
 cosmic microwave background and the large-scale structure of the
 universe. The ``$\delta N$ formalism'' is a popular and powerful
 technique for computing nonlinear evolution of cosmological
 perturbations on large scales. In particular, it enables us to compute
 the curvature perturbation, $\zeta$, on large scales without actually
 solving perturbed field equations. However, people often wonder why
 this is the case. In order for this approach to be valid,
 the perturbed Hamiltonian constraint and matter-field equations on
 large scales must, with a suitable choice of coordinates, take on the
 same forms as the corresponding unperturbed equations. We find that
 this is possible when (1) the unperturbed metric is given by a
 homogeneous and isotropic Friedmann-Lema\^itre-Robertson-Walker metric;
 and (2) on large scales and with a suitable choice of coordinates, one
 can ignore the shift vector ($g_{0i}$) as well as time dependence of
 tensor perturbations to $g_{ij}/a^2(t)$ of the perturbed metric. While the
 first condition has to be assumed {\it a priori}, the second condition
 can be met when (3) the anisotropic stress becomes negligible on large
 scales. However, in order to explicitly show that the second condition
 follows from the third condition, one has to use gravitational field
 equations, and thus this statement may depend on the details of 
 the theory of gravitation. Finally, as the $\delta N$ formalism uses only
 the Hamiltonian constraint and matter-field equations, it does not {\it
 a priori} respect the momentum constraint. We show that the
 error in the momentum constraint only yields a decaying
 mode solution for 
 $\zeta$, and the error vanishes when the slow-roll conditions are
 satisfied.
\end{abstract} 
\preprint{TCC-009-12}
\maketitle
\section{Introduction}
Given the success of cosmological linear perturbation theory, the focus
has shifted to {\it nonlinear} evolution of cosmological
perturbations. As the magnitude of the primordial curvature perturbation
is of order $10^{-5}$, any nonlinearities are expected to be small;
however, such nonlinearities can be measured using non-Gaussian
correlations of cosmological perturbations (such as temperature and
polarization anisotropy of the cosmic microwave background
\cite{Komatsu:2010} and density fluctuations in the large-scale
structure of the universe \cite{Desjacques/Seljak:2010}). For this
reason, precise understanding of the nonlinear evolution of
cosmological perturbations is of great interest in cosmology. 

The so-called ``$\delta N$ formalism'' 
\cite{Starobinsky:1982,*Starobinsky:1986,Salopek/Bond:1990,Sasaki/Stewart:1996,David/etal:2005,Lyth/Rodriguez:2005}
is a popular technique for computing non-linear evolution of
cosmological perturbations on large scales. Here, by ``large scales,''
we mean the scales greater than the Hubble horizon, in a sense that the
comoving wavenumber of perturbations $k$ is much less than the
reciprocal of the comoving Hubble length, i.e., $k\ll aH$. In
particular, it enables us to compute the curvature perturbation
$\zeta$ without actually solving the perturbed field equations. 
In this paper, we show why this is the case by rederiving the $\delta N$
formalism using the gradient expansion method as applied to Einstein's
field equations and scalar-field equations in the flat gauge. 
The usual derivation of the $\delta N$
formalism is based on the so-called ``separate universe''
approach \cite{Wands/etal:2000}, which assumes the existence of a locally 
homogeneous (but not necessarily isotropic \cite{Dimopoulos/etal:2009})
region smoothed over some large length scale.
We provide a support for this assumption by considering a
global region including many such smoothed local regions
and show that 
they behave as if they were locally homogeneous regions which evolve
independently from each other.
In so doing, we point out a subtlety regarding the momentum constraint, which
is not {\it a priori} respected by the $\delta N$ formalism. 

The organization of this paper is as follows. In Sec. II, we describe
our basic setup including the metric, gauge, and scalar-field
Lagrangian. In Sec. III, we review the gradient expansion method, which
constitutes the basis for the $\delta N$ formalism. In Sec. IV, we
rederive the $\delta N$ formalism. In Sec. V, we give the sufficient
conditions for the validity of the $\delta N$ formalism and conclude. 

\section{Basic Setup}
\subsection{Metric}
We write the spacetime metric in the Arnowitt-Deser-Misner (ADM) form, which is the standard (3+1)-decomposition of the metric \cite{ADM:1962}
\begin{equation}
		ds^2 = - \alpha^2 dt^2 + \gamma_{ij} \left( dx^i + \beta^i dt \right)\left( dx^j + \beta^j dt \right) ,
	\label{}
\end{equation}
where $\gamma_{ij}$ is decomposed as follows:
\begin{equation}
		\gamma_{ij} \equiv a^2 e^{2\psi} \left( e^h \right)_{ij}.
		\label{}
\end{equation}
Here, $a$ is the scale factor which depends only on time, and $\psi$ is the scalar perturbation to the spatial curvature.  
The traceless tensor $h_{ij}$ is further decomposed as 
\begin{equation}
		h_{ij} = \partial_i C_j + \partial_j C_i - \frac{2}{3} \delta_{ij} \partial_k C^k + h^{(T)}_{ij},
		\label{}
\end{equation}
where $C_i$ contains both scalar and vector perturbations, whereas $h_{ij}^{(T)}$ represents tensor perturbations. 

We decompose the extrinsic curvature $K_{ij}$ into a trace part $K$ and a traceless part $\tilde{A}_{ij}$ as\footnote{In Salopek and Bond~\cite{Salopek/Bond:1990}, a trace-free part of the extrinsic curvature is denoted as $\bar{K}_{ij}$. In our notation, we have $\bar{K}_{ij}=a^2 e^{2\psi} \tilde{A}_{ij}$.}
\begin{equation}
 K_{ij} =\frac{\gamma_{ij}}{3}K + a^2e^{2\psi}\tilde{A}_{ij}.
\end{equation}
Einstein's field equations written in terms of these variables are summarized in Appendix A.

\subsection{Flat gauge}
In this paper, we shall fix the gauge completely (i.e., leaving no gauge degree of freedom) by imposing the following gauge-fixing condition\footnote{Sometimes another gauge condition $\dot{\psi} = 0$ and $\beta_i = 0$ is imposed; however, the degree of freedom of this gauge is not completely fixed, as this gauge condition only gives the relation for the time derivatives of the shifts in temporal and spatial coordinates.}:
\begin{equation}
 \psi = C_i = 0.
\label{gauge-fixing}
\end{equation}
Therefore, the spatial metric is described only by the scale factor and tensor perturbations as $\gamma_{ij} = a^2 [e^{h^{(T)}}]_{ij}$. This gauge was also used by \cite{Maldacena:2003} [see his Eq.~(3.2)].

We shall call this gauge the ``flat gauge'' throughout this paper. Note that a flat-gauge condition in the literature sometimes does not include
$C_i=0$. In such a case a residual gauge degree of freedom would remain. In the flat gauge, the metric is given by
\begin{equation}
		ds^2 = - \alpha^2 dt^2 + a^2(t) [e^{h^{(T)}}]_{ij} \left( dx^i + \beta^i dt \right)\left( dx^j + \beta^j dt \right).
	\label{}
\end{equation}
The variables in this metric such as $\alpha$, $\beta_i$, and
$h^{(T)}_{ij}$ contain {\it nonlinear} perturbations. However, we shall
assume that the unperturbed metric is still given by a homogeneous and isotropic Friedmann-Lema\^itre-Robertson-Walker metric:
\begin{equation}
		ds^2 = - dt^2 + a^2(t) \delta_{ij} dx^idx^j.
	\label{background}
\end{equation}
Therefore, our argument below does not hold if the unperturbed metric is not given by Eq.~(\ref{background}).

\subsection{Scalar-field Lagrangian}
We shall consider a universe filled with scalar fields:
\begin{equation}
		\mathcal{L} = - \frac{1}{2} G_{IJ}\partial^{\mu} \varphi^I \partial_{\mu} \varphi^J - V .
		\label{}
\end{equation}
The capital latin indices ($I$, $J$, etc.) denote scalar-field components running from 1 to $n$ where $n$ is the number of scalar fields. Here, $G_{IJ}$ is the metric tensor for scalar-field space. For simplicity, we shall take the canonical kinetic term $G_{IJ} = \delta_{IJ}$ for the moment. We then argue later (in Sec.~\ref{sec:noncanonical}) that the results are also valid for noncanonical kinetic terms in the first order of the gradient expansion.

With this Lagrangian, the stress-energy tensor and the field equation of
scalar fields are given by 
\begin{equation}
		T_{\mu\nu} =G_{IJ} \partial_{\mu}\varphi^I \partial_{\nu} \varphi^J
		+ g_{\mu\nu} \left( - G_{IJ} \frac{1}{2} \partial^{\alpha} \varphi^I \partial_{\alpha} \varphi^J  - V\right),
		\label{}
\end{equation}
\begin{equation}
		\frac{1}{\sqrt{g}} \partial_{\mu} \left(\sqrt{g} g^{\mu\nu} \partial_{\nu} \varphi^I \right) - V_I = 0,
		\label{motion}
\end{equation}
where $V_I \equiv \partial V/ \partial \varphi^I$.

\section{Gradient expansion method}
\subsection{Ordering in the gradient expansion}
\label{sec:estimation}

Since we are interested in nonlinear perturbations on superhorizon scales,
we shall expand field equations in the number of spatial derivatives:
this is called the gradient expansion method
\cite{Salopek/Bond:1990,David/etal:2005}. In this method, the ratio of the comoving wavenumber and the comoving Hubble scale, 
\begin{equation}
\epsilon \equiv \frac{k}{aH}
\end{equation} 
is taken to be a small parameter. 

Before we proceed, let us emphasize that we assume
the validity of perturbative expansion: namely, while we shall deal with
nonlinear perturbations, we assume that the $(i+1)$th-order
perturbations are smaller 
than the $i$th-order perturbations. This means that we have
{\it two} smallness parameters: one is the number of derivatives,
$\epsilon$, and the other is a smallness parameter of perturbation
theory $\delta$, which corresponds to $\psi$, $\beta^i$,
$\varphi^I-\bar{\varphi}^I$, etc. 

These two parameters should satisfy the following condition:
\begin{equation}
		\delta < \epsilon \ll 1.
		\label{condition}
\end{equation}
This is because, if we take the smoothing length to be infinitely large,
i.e., $\epsilon\to 0$, then the perturbation must vanish, i.e.,
$\delta\to 0$. Then, the metric must approach the unperturbed metric
given by Eq.~(\ref{background}) as we take $\epsilon\to 0$. In other
words, the amplitude of the perturbations should be limited by the
smoothing length we take.

We now estimate the ordering of perturbation variables in
terms of the gradient expansion.
First, we demand that all physical quantities do not vanish in the lowest order of the gradient expansion:
\begin{align}
		&\alpha-1  = \tilde{\gamma}_{ij} - \delta_{ij} = \varphi^I - \bar{\varphi}^I = \mathcal{O}(\epsilon^0,\delta), \notag \\
 &\hspace{1cm} \beta^i = {\cal O}(\epsilon^{-1},\delta).
 \label{first-estimation}
\end{align}
We do not include $\psi$ here because we work in the flat gauge.
We have defined $\tilde{\gamma}_{ij}\equiv
\gamma_{ij}/[a^2(t)e^{2\psi}]$, which is equal to $[e^{h^{(T)}}]_{ij}$ in the
flat gauge.

One may wonder why we chose to start with $\beta^i={\cal
O}(\epsilon^{-1})$, which seems to diverge in the limit of $\epsilon\to
0$. However, this is not true.
As noted earlier, the existence of the perturbation ($\delta > 0$) guarantees $\epsilon > 0$ 
and we always have $\beta^i = {\cal O}(\epsilon^{-1},\delta) < 1$ from Eq.~(\ref{condition}); thus, there is no divergence in the metric.
In fact, we recover the standard Friedmann equation in the lowest order
approximation. Furthermore, at the end of Sec.~\ref{sec:momentum}, we
show that consistency 
between the Hamiltonian and momentum constraint equations demands
$\beta^i={\cal O}(\epsilon^{-1})$. 

Note that the shift vector 
comes with a spatial derivative $\partial_i$ in Einstein's field equations and scalar-field equations. 
As $\partial_i\beta^i={\cal O}(\epsilon^0)$, the spatial derivatives are
kept in Einstein's field equations and scalar-field equations 
for $\alpha-1 = \tilde{\gamma}_{ij} - \delta_{ij} = \varphi^I -
\bar{\varphi}^I = {\cal O}(\epsilon^0)$.
In other words, as we keep spatial derivatives 
in our approach, we are considering some
global region 
in which there are many smoothed local regions. Therefore, we do not
{\it a priori} demand that these local regions evolve independently of
each other, contrary to what is always demanded by a separate universe
approach. Specifically, for a separate universe approach, $\beta^i =
{\cal O}(\epsilon)$ is always assumed {\it a priori}.

Similarly, when we decompose the quantities $\beta_i$ and $C_i$ into scalar and vector components as $\beta_i = \partial_i \beta^{(S)} + \beta_i^{(V)}$ and $C_i = \partial_i C^{(S)} + C_i^{(V)}$, respectively, the scalar components are of order $\epsilon^{-2}$: $\beta^{(S)} = C^{(S)} = {\cal O}(\epsilon^{-2})$. 

In order to see how Eq.~(\ref{first-estimation}) can be relaxed, we now investigate the nature of solutions for $\beta_i$ and $h^{(T)}_{ij}$. 

At the first order in perturbation variables and the lowest order in the gradient expansion,
the evolution equation for $\tilde{A}_{ij}$ is given by [see Eq.~(\ref{full-A-evolusion})]
\begin{equation}
		\dot{\tilde{A}}_{ij} + 3H \tilde{A}_{ij} = {\cal O}(\epsilon,\delta^2),
		\label{0th-A}
\end{equation}
where $H$ is the Hubble expansion rate $H \equiv
\dot{a}/a$. Here, we have ignored the anisotropic stress
term on the right-hand side of Eq.~(\ref{full-A-evolusion}), as it is of
the second order in the gradient expansion for scalar fields. This is a
stronger-than-necessary condition: Eq.~(\ref{0th-A}) is still valid if
the anisotropic stress of matter fields is of the first order in the
gradient expansion. 

It follows from Eq.~(\ref{0th-A}) that the traceless part of the
extrinsic curvature $\tilde{A}_{ij}$ has a decaying solution
$\tilde{A}_{ij}\propto 1/a^3$
\cite{Salopek/Bond:1990}.\footnote{
As we start with $\beta^i={\cal O}(\epsilon^{-1})$, we need to linearize
Eq.~(\ref{full-A-evolusion}) to obtain Eq.~(\ref{0th-A}), showing
$\tilde{A}_{ij}\propto 1/a^3$. On the other hand, assuming $\beta^i={\cal
O}(\epsilon)$, Hamazaki derives
$\tilde{A}_{ij}\tilde{A}^{ij}\propto 1/a^6$ without using perturbation
theory (see Eq.~(2.54) of \cite{Hamazaki:2008}).} On the other hand, the evolution equation for $\gamma_{ij}$ with $C_i=0$ yields [see Eq.~(\ref{full-gamma})] 
\begin{equation}
		\dot{h}^{(T)}_{ij} = -2 \tilde{A}_{ij} + \frac{1}{a^2}
		\left( \partial_i \beta_j + \partial_j \beta_i - \frac{2}{3} \delta_{ij} \delta^{kl} \partial_k \beta_l \right).
		\label{0th-gamma}
\end{equation}
As the scalar, vector, and tensor modes are independent in
linear theory, the equations for the shift vector and tensor
perturbations are given by 
\begin{eqnarray}
		\ddot{h}_{ij}^{(T)} + 3 H \dot{h}_{ij}^{(T)} = {\cal O}(\epsilon,\delta^2), \nonumber \\
		\dot{\beta}^i + 3H \beta^i = {\cal O}(\epsilon^0,\delta^2) .
		\label{h_beta}
\end{eqnarray}
Therefore, $\dot{h}^{(T)}_{ij}$ and $\beta^i=\beta_i/a^2$ also have decaying solutions scaling as $a^{-3}$. This is a consequence of the fact that the unperturbed metric [Eq.~(\ref{background})] is given by a homogeneous and isotropic Friedmann-Lema\^itre-Robertson-Walker metric. In other words, this result may not hold for anisotropic models such as Bianchi-type metrics.

At the second order in perturbation variables, 
as the source terms in Eqs.~(\ref{full-gamma}) and (\ref{full-A-evolusion}) are decaying, the second-order equations for $h^{(T)}_{ij}$ and $\beta_i$ have approximately the same forms as the first-order equations [Eqs.~(\ref{h_beta}) and (\ref{0th-A})], 
and thus their solutions must also be decaying as $a^{-3}$. Similarly,
$n$th-order solutions for $n\ge 3$ are also decaying.

These properties allow us to safely ignore, in the lowest order of the
gradient expansion and the $n$th order of perturbation theory, the traceless part of the extrinsic curvature
$\tilde{A}_{ij}$, the shift vector $\beta^i$, as well as a time
derivative of tensor perturbations $\dot{h}_{ij}^{(T)}$, after the
decaying solutions become sufficiently small. This means that these quantities must be higher order in the gradient expansion than naively assumed in Eq.~(\ref{first-estimation}): $\tilde{A}_{ij} = {\cal O}(\epsilon)$, $\beta^i = {\cal O}(\epsilon^0)$, and $\dot{h}^{(T)}_{ij} = {\cal O}(\epsilon)$.

Now, it turns out that the above argument also applies to the
next order of the gradient expansion. At the next order in the gradient
expansion
\begin{eqnarray}
	&&\alpha-1 = \tilde{\gamma}_{ij}-\delta_{ij} = \varphi^I - \bar{\varphi}^I = {\cal O}(\epsilon), \nonumber \\
	&& \hspace{1.7cm}\beta^i = {\cal O}(\epsilon^0),
\end{eqnarray}
one can show that for scalar fields whose anisotropic stress is of the
second order in the gradient expansion, the equations take on the same
form as Eqs.~(\ref{0th-A}) and (\ref{h_beta}):
\begin{eqnarray}
\label{aijfull}
		\dot{\tilde{A}}_{ij} + 3H \tilde{A}_{ij} &=&  {\cal O}(\epsilon^2,\delta^2), \\
		\ddot{h}_{ij}^{(T)} + 3H \dot{h}_{ij}^{(T)} &=&  {\cal O}(\epsilon^2,\delta^2), \\
		\dot{\beta}^i + 3 H \beta^i &=&  {\cal O}(\epsilon,\delta^2).
\end{eqnarray}
Applying the same
argument as above, one finds that $\tilde{A}_{ij}$,
$\dot{h}_{ij}^{(T)}$, as well as $\beta^i$ decay for the $n$th order in
perturbation theory. 
We thus find:
$\tilde{A}_{ij} = {\cal O}(\epsilon^2)$, $\beta^i = {\cal O}(\epsilon)$,
and $\dot{h}^{(T)}_{ij} = {\cal O}(\epsilon^2)$. However, this argument
cannot be extended to the second order of the gradient expansion, as
Eq.~(\ref{aijfull}) is valid only when the anisotropic stress term is
unimportant. As one can no longer ignore the anisotropic stress of
scalar fields at the second order in the gradient expansion,
Eq.~(\ref{aijfull}) is no 
longer valid in that order.

Therefore, Eq.~(\ref{first-estimation}) should be revised as
\begin{align}
		&  \hspace{1cm}  \alpha-1 = \varphi^I- \bar{\varphi}^I = {\cal O}(\epsilon^0), \notag \\
        &  \beta^i = {\cal O}(\epsilon), \ \ \ \dot{h}^{(T)}_{ij} = {\cal O}(\epsilon^2),
		\label{estimation}
\end{align}
where we have dropped $\delta$ in ${\cal O}(\dots)$, as
the above estimation is valid for all orders of perturbation theory.
Note that this result is valid only in the flat gauge given by Eq.~(\ref{gauge-fixing}). In particular, the condition $C_i=0$ was needed to estimate the gradient-expansion order of $\beta^i$ and $\dot{h}^{(T)}_{ij}$.

These results might depend on the details of the theory of gravitation,
as we have used Einstein's field equations to obtain solutions of
$\beta^i$ and $\dot{h}^{(T)}_{ij}$. Furthermore, as a perturbative
expansion is used in estimating $\beta^i$ and $\dot{h}^{(T)}_{ij}$, the
validity of a perturbative description of the metric with the
unperturbed metric given by Eq.~(\ref{background}) has been assumed in
the above argument. 

\subsection{Comparison with previous work}
\label{sec:previous}
How does Eq.~(\ref{estimation}) compare with the previous
work? Our starting point, Eq.~(\ref{first-estimation}), is different
from the assumption made in Lyth, Malik and Sasaki
\cite{David/etal:2005} (also see \cite{Lyth/Liddle:2009}).
They assume that 
there exists an approximate set of coordinates with which the metric of
any local region can be written as a
Friedmann-Lema\^itre-Robertson-Walker metric. This implies that the
shift vector $\beta_i$ vanishes and the quantity
$\tilde{\gamma}_{ij}$ is time independent in the limit of $\epsilon \to
0$: $\beta_i = {\cal O}(\epsilon)$ and $\dot{\tilde{\gamma}}_{ij}={\cal
O}(\epsilon)$. We do not make this assumption {\it a priori}, and thus
our argument is more general than that given in
\cite{David/etal:2005}. They then show that by using Einstein's field
equations and ignoring the anisotropic stress term,
$\dot{\tilde{\gamma}}_{ij}$ decays in the first order of the gradient
expansion, concluding that $\dot{\tilde{\gamma}}_{ij}={\cal O}(\epsilon^2)$.

In \cite{Weinberg1,*Weinberg2}, Weinberg uses a broken
symmetry argument to show  $\beta_i = {\cal O}(a^{-2})$
and $\dot{\tilde{\gamma}}_{ij}={\cal
O}(a^{-2})$ for generally covariant theories and
with a suitable choice of coordinates, assuming that the unperturbed
metric is given by Eq.~(\ref{background}) and the anisotropic stress
term is negligible. He then shows that for Einstein's field equations
in a coordinate system in which $\beta_i=0$ and a certain combination of
matter perturbations vanishes, this solution is an attractor. By
identifying ${\cal O}(a^{-2})$ with ${\cal O}(\epsilon^2)$ (because each
spatial derivative must come with $1/a$), his argument yields  $\beta_i = {\cal O}(\epsilon^{2})$
and $\dot{\tilde{\gamma}}_{ij}={\cal
O}(\epsilon^{2})$.

Therefore, our finding agrees with the previous work: if
the unperturbed metric is given by Eq.~(\ref{background}), the
anisotropic stress term is negligible on large scales, and if field
equations are given by Einstein's field equations, then $\beta_i = {\cal
O}(\epsilon)$ and $\dot{\tilde{\gamma}}_{ij}={\cal O}(\epsilon^2)$. Note
that Weinberg's estimate for the order of $\beta_i$ is higher by $\epsilon$;
however, he does not show $\beta_i = {\cal O}(\epsilon^2)$
explicitly because he chooses coordinates in which $\beta_i=0$.

\subsection{Gradient expansion of Einstein's field equations and scalar-field equations in the flat gauge}
\label{sec:noncanonical}
Now, we apply the gradient expansion to Einstein's field equations and
scalar-field equations. We shall work with the flat gauge given by
Eq.~(\ref{gauge-fixing}), which gives the gradient-expansion order of
perturbation variables given in Eq.~(\ref{estimation}). 

We shall choose the number of $e$-folds $N \equiv \int_{t_\ast}^t
Hdt^{\prime}$ as our time coordinates in the flat gauge. The
Hamiltonian constraint 
[Eq.~(\ref{full_hami})] and the scalar-field equation
[Eq.~(\ref{motion})] in {\it both} the lowest order and the next order of the gradient expansion are given by
\begin{equation}	
		3\tilde{H}^2M_p^2  = \rho  ,
		\label{0th-hami}
\end{equation}
\begin{equation}
		\tilde{H} \partial_N \left( \tilde{H} \varphi^I_N \right) + 3 \tilde{H}^2 \varphi^I_N + V_I = 0 ,
		\label{0th-motion-2}
\end{equation}
where $\tilde{H} \equiv H/\alpha$ is related to a trace of the extrinsic curvature as $\tilde{H}=-K/3$, and the subscript $N$ denotes a partial derivative with respect to $N$. The energy density $\rho$ is given by $\rho = \frac{\tilde{H}^2}{2} G_{IJ} \varphi_N^I \varphi_N^J + V$. 
Then, $\tilde{H}$ is given by \cite{Sasaki:1998}
\begin{equation}
		\tilde{H}^2 = \frac{2V}{6M_p^2 - G_{IJ} \varphi^I_N \varphi^J_N}.
		\label{0th-hami-2}
\end{equation}
All we need to do is to solve Eq.~(\ref{0th-motion-2}) coupled with Eq.~(\ref{0th-hami-2}).

On the other hand, the unperturbed equations are
\begin{equation}
		3H^2M_p^2  = \bar{\rho} ,
		\label{back-hami}
\end{equation}
\begin{equation}
		H \partial_N \left( H \bar{\varphi}^I_N \right) + 3H^2 \bar{\varphi}^I_N + V_I(\bar{\varphi}) = 0 ,
		\label{back-motion}
\end{equation}
where $\bar{\rho}$ and $\bar{\varphi}^I$ are the unperturbed energy density and scalar fields, respectively.
Apparently, the perturbative equations [Eqs.~(\ref{0th-hami}) and
(\ref{0th-motion-2})] coincide exactly with the unperturbed equations
[Eqs.~(\ref{back-hami}) and (\ref{back-motion})]. 
This result shows that each region smoothed by a superhorizon scale $\epsilon \ll 1$
in the universe evolves independently and behaves like an unperturbed
universe providing a support for the assumption made by a separate
universe approach.

These results might depend on the details of the theory of gravitation.
While the correspondence between the perturbed and
unperturbed equations for other theories of gravitation is an interesting
problem, in this paper we shall focus on Einstein's General
Relativity. However, these results should not depend on the form of the
Lagrangian of scalar fields. This is because the anisotropic
stress (i.e., a traceless part of the stress-energy tensor) for scalar
fields with arbitrary Lagrangian necessarily comes with two spatial
derivatives, and thus it must be ${\cal O}(\epsilon^2)$. As a result,
$\tilde{A}_{ij}$ has a decaying solution and a
Friedmann-Lema\^itre-Robertson-Walker universe will be restored on large
scales.

One can generalize the above results for the canonical case to noncanonical Lagrangians given by $\mathcal{L} = P(X^{IJ},\varphi^K)$, where $X^{IJ} \equiv -g^{\mu\nu} \partial_{\mu} \varphi^I\partial_{\nu} \varphi^J$. The Hamiltonian constraint is still given by Eq.~(\ref{0th-hami}) while the scalar-field equation is given by
\begin{equation}
\tilde{H} \partial_N \left[ \tilde{H} \partial_N \varphi^J_N \right] P_{IJ}
		+ \tilde{H}^2 \varphi^J_N \partial_N P_{IJ} + 3\tilde{H}^2 P_{IJ} \varphi^J_N + \frac{P_I}{2} = 0,
\end{equation}
where $P_{IJ} \equiv \partial P/\partial X^{IJ}$, $X^{IJ} = \tilde{H}^2\varphi^I_N  \varphi^J_N$, and the energy density is defined as $\rho = 2X^{IJ} P_{IJ} - P$. As $P$ and $P_{IJ}$ are functions of $X^{IJ}$, $\varphi^K$, and $\tilde{H}$, one can write $\tilde{H}$ as a function of $X^{IJ}$ and $\varphi^K$ if an explicit form of the Lagrangian $P$ is specified. 

What are the implications of these results? As the equations take on the same forms, the functional forms of the solutions for the perturbed equations and those for the unperturbed equations must be the same. Therefore, the  perturbed solutions $\varphi^I$ are given by the unperturbed solutions $\bar{\varphi}^I$ {\it with perturbed initial conditions computed in the flat gauge, $\varphi^I_{\ast}(\xx)$ and $\varphi^I_{N\ast}(\xx)$:}
\begin{equation}
		\varphi^I(N,\xx) = \bar{\varphi}^I(N,\varphi^J_{\ast}(\xx),\varphi^K_{N\ast}(\xx)).
		\label{full-solution}
\end{equation}
{\it This is the fundamental result of the gradient expansion as applied to Einstein's field equations and scalar-field equations in the flat gauge.} Here, the subscript $*$ indicates that the quantity is evaluated at some initial time, where all the relevant fields are sufficiently outside their sound horizon, i.e., $k\ll a(t_*)H(t_*)/c_s^I$, where $c_s^I$ is the speed of sound of propagation of an $I$th scalar-field perturbation. 

In order to simplify our notations, from now on we shall use the lower-case alphabet indices, such as $a, b, c \dots$, to denote the numbers of scalar fields and their time derivatives: $$\varphi^a \equiv (\varphi^I,\varphi^J_N),$$ with $a$ running from 1 to $2n$. With this notation, the solution [Eq.~(\ref{full-solution})] is expressed as $\varphi^a(N,\xx) = \bar{\varphi}^a(N,\varphi_{\ast}^b(\xx))$. 

Similarly, we can write the perturbed energy density of multiscalar fields using the unperturbed energy density solution $\rho(N,\xx) = \bar{\rho}(N,\varphi^a_{\ast}(\xx))$.

\section{The $\delta N$ formalism}
We now need to relate perturbed initial scalar fields and their derivatives $\varphi^a_*(\xx)$ to the observables. In cosmology, it is now customary to express the observables such as temperature and polarization anisotropies and the large-scale distribution of galaxies in terms of a curvature perturbation in the ``uniform-density gauge,'' denoted as $\zeta$.

The so-called $\delta N$ formalism \cite{Starobinsky:1982,*Starobinsky:1986,Salopek/Bond:1990,Sasaki/Stewart:1996,David/etal:2005,Lyth/Rodriguez:2005} achieves this by realizing that $\zeta$ is equal to a perturbation to the number of $e$-folds, $N$, arising from perturbed initial scalar fields $\varphi^a_*(\xx)$ computed in the flat gauge. 

\subsection{Conservation of $\zeta$ outside the horizon}
We define  the uniform-density gauge as $\delta \rho = 0$ and $C_i=0$. (Once again, a uniform-density gauge in the literature sometimes does not include $C_i=0$. In such a case the gauge is not completely fixed.) Let us denote a value of $\psi$ in the uniform-density gauge as $\psi|_{\delta \rho = C_i = 0}$\footnote{Incidentally, as the gauge is completely fixed for $\psi|_{\delta \rho = C_i =0}$, there is no ambiguity with respect to the residual gauge degree of freedom. Moreover, one can always write perturbation variables (such as $\psi$) after gauge fixing as a combination of perturbation variables before gauge fixing (such as $C_i$ and $\delta \rho$) such that $\psi$ is explicitly gauge-invariant. For example, we have, at the linear order,
\begin{equation}
	\zeta \equiv \psi|_{\delta \rho = C_i =0}
		= \psi - \frac{\partial_i C^i}{3}- \frac{\delta \rho}{\bar{\rho}_N} ,
		\label{ex}
\end{equation}
where the right-hand side of Eq.~(\ref{ex}) is the well-known form for a gauge-invariant curvature perturbation in the linear order \cite{Bardeen/Turner/Steinhardt:1983}.}, and write $\zeta$ as 
\begin{equation}
 \zeta \equiv \psi|_{\delta \rho = C_i =0}.
\end{equation}

This quantity is useful for extracting information about the physics of
inflation, as it is conserved outside the horizon, provided that the
adiabatic condition $p= p[\rho]$ is satisfied
\cite{Wands/etal:2000,David/etal:2005}. This is easily seen from the
energy conservation 
equation in the lowest order as well as in the next order gradient expansion with the gauge condition $C_i = 0$:
\begin{equation}
		\dot{\rho} + 3 (H + \dot{\psi}|_{C_i=0})(\rho + p) = 0.
		\label{rho}
\end{equation}
The perturbation to this equation in the uniform-density gauge yields
\begin{equation}
		\dot{\psi}|_{\delta \rho = C_i = 0} = 0,
\end{equation} 
and thus $\zeta=\psi|_{\delta \rho = C_i = 0}$ becomes a constant, provided that the adiabatic condition is satisfied.  

Alternatively, Eq.~(\ref{rho}) may be integrated with respect to $t$ without imposing $\delta\rho=0$: 
\begin{equation}
		\tilde{\zeta} \equiv \psi + \int_{\bar{\rho}}^{\rho} \frac{d \rho}{3 (\rho + p[\rho])}  = \mbox{const.} 
		\label{conservezeta}
\end{equation}
One may then identify this quantity $\tilde{\zeta}$ as a generalization of $\zeta$ when $\delta\rho=0$  is not imposed; however, $\tilde{\zeta}$ is not gauge invariant and does not coincide with $\zeta = \psi|_{\delta \rho = C_i = 0}$, unless the adiabatic condition is satisfied.

What about scalar fields? As scalar fields do not satisfy the adiabatic
condition in general, $\zeta$ is not conserved in a universe filled with
scalar fields. However, as shown by \cite{Weinberg:2003}, $\zeta$ is
generally conserved outside the horizon when inflation was driven by a
single scalar field. More precisely, $\zeta$ is conserved outside the
horizon  in a universe dominated by a single scalar field, provided that
the slow-roll conditions are satisfied, or that we completely neglect a
decaying mode solution without imposing the slow-roll conditions. This
implies that the slow-roll conditions correspond effectively to the
adiabatic condition and the neglect of a decaying mode solution
for a single scalar field.

\subsection{Relation between $\zeta$  and the difference in the number of $e$-folds} 
The relation between the curvature perturbation and the number of
$e$-folds is given by the gauge transformation of the spatial metric
$\gamma_{ij}$. Under a gauge transformation given by $t\to T=t+\delta T$
and $x^i\to X^i=x^i+\xi^i$, the metric transforms as
$g_{ij}(t,\xx)\to \hat{g}_{ij}(T,\mbox{\boldmath $X$})$. 
Let us write the 3-metric in the original coordinates in terms of the
3-metric in the new coordinates:
\begin{align}
		\gamma_{ij}(t,\xx)
		= &  -\hat{\alpha}^2(T,\mbox{\boldmath $X$}) \frac{\partial\delta
 T}{\partial x^i}\frac{\partial \delta T}{\partial x^j} \notag \\
		& + \hat{\beta}_k(T,\mbox{\boldmath $X$}) \frac{\partial
 X^k}{\partial x^i} \frac{\partial \delta T}{\partial x^j}
		+ \hat{\beta}_k(T,\mbox{\boldmath $X$}) \frac{\partial X^k}{\partial
 x^j}\frac{\partial \delta T}{\partial x^i} \notag \\
	    & + \hat{\gamma}_{kl}(T,\mbox{\boldmath $X$})\frac{\partial X^k}{\partial x^i} \frac{\partial X^l}{\partial x^j}.
		\label{gauge}
\end{align}
We shall always impose $C_i=0$, which completely fixes the spatial gauge degree of freedom, and thus we can set $\xi^i=0$ without loss of generality.

Let us examine each term in terms of the gradient-expansion order. The first term is of order ${\cal O}(\epsilon^2)$.
As shown in Sec.~\ref{sec:estimation}, when $C_i = 0$, the shift vector is of ${\cal O}(\epsilon)$; thus, the second and third terms are of order ${\cal O}(\epsilon^2)$. This means that, up to ${\cal O}(\epsilon^2)$, the 3-metric transforms as
\begin{equation}
		{\gamma}_{ij}|_{C_i = 0}(t,\xx) = \hat{\gamma}_{ij}|_{C_i =0}(T,\xx) + {\cal O}(\epsilon^2).
		\label{gammatrans}
\end{equation}

Recalling $\gamma_{ij}=a^2(t)e^{2\psi}(e^h)_{ij}$ and taking the
determinant and logarithm of both sides of Eq.~(\ref{gammatrans}), we find
\begin{equation}
		\psi|_{C_i = 0}(t,\xx) = \hat{\psi}|_{C_i =
		 0}(T,\xx) + \ln\left( \frac{a(T)}{a(t)} \right) . 
		\label{psi_C}
\end{equation}
Thus, $\psi|_{C_i = 0}$ approximately transforms as a scalar quantity having $\ln (a)$ as the unperturbed value.
It follows from Eq.~(\ref{psi_C}) that the gauge transformation of $\psi|_{C_i = 0}$ from the flat gauge (in which $\psi=0$) into the uniform-density gauge is given by
\begin{equation}
		\psi|_{\delta \rho = C_i = 0}(T,\xx) 
= \ln \left(\frac{a(t)}{a(T)}\right),
		\label{1-definition-deltaN}
\end{equation}
where $T$ denotes time coordinates in the uniform-density
gauge. 

On the other hand, when we go from the flat gauge to the uniform-density gauge, the number of $e$-folds, $N \equiv \int_{t_*}^t H dt'$, transforms as $N \to \hat{N}$, where
\begin{equation}
		\hat{N} \equiv \int^{T}_{t_{\ast}} H(t^{\prime}) dt^{\prime} = \ln \left( \frac{a(T)}{a(t_{\ast})} \right) .
		\label{Nhat}
\end{equation}
Here, $t_*$ is an arbitrary initial time.

Comparing Eq.~(\ref{1-definition-deltaN}) to
Eq.~(\ref{Nhat}), one finds
\begin{align}
 \psi|_{\delta \rho = C_i = 0}(T,\xx) & = \ln \left( \frac{a(t)}{a(t_*)} \right) - \ln \left( \frac{a(T)}{a(t_*)} \right) \notag \\
		& = N - \hat{N} \equiv \delta N.
\end{align}
Therefore, $\zeta=\psi|_{\delta \rho = C_i = 0}$ is equal to the
difference between 
the number of $e$-folds computed in the flat gauge and that computed in the
uniform-density gauge. 
The remaining task is to relate $\delta N$ to perturbed
initial scalar fields in the flat gauge.

\subsection{Relation between $\delta N$  and perturbed initial scalar fields in the flat gauge}
The most important result that came from the gradient expansion of
Einstein's field equations and scalar-field equations in the flat gauge
is that perturbed quantities can be calculated using their unperturbed
solutions with perturbed initial scalar-field values and their time
derivatives computed in the flat gauge. Therefore, a perturbed energy
density in the flat gauge is given by $\rho(N,\xx) =
\bar{\rho}(N,\varphi_*^a(\xx))$. Here, we choose the number of $e$-folds
as time coordinates. 

On the other hand, by definition the energy density in the
uniform-density gauge (whose time coordinates are denoted as $\hat{N}$) 
is equal to the unperturbed density. Namely, when we go from the flat
gauge to the uniform-density gauge by changing the time coordinates as
$N\to \hat{N}= N + \delta N$, the density transforms as 
$\rho(N,\xx)\to
\hat{\rho}(\hat{N},\xx)=\bar{\rho}(\hat{N})$.
Here, $C_i = 0$ is satisfied in both gauges, and thus there is no
ambiguity with respect to the spatial gauge degree of freedom. 
Now, as the energy density is a four scalar,  
\begin{equation}
		\rho(N,\xx) = \hat{\rho}(\hat{N},\xx)=\bar{\rho}(\hat{N}),
		\label{gauge2}
\end{equation}
which gives $\bar{\rho}(N,\varphi_*^a(\xx))=\bar{\rho}(\hat{N})$. Inverting this result yields
\begin{equation}
 N=\hat{N}(\bar{\rho},\bar{\varphi}_*^a(\xx)),
\end{equation}
where the functional form of $\hat{N}$ is the same as that
of the unperturbed 
number of $e$-folds. That the unperturbed density $\bar{\rho}$ (not $t$ or $N$) is used as the time coordinates here ensures that the final time slice coincides with the uniform density hypersurface.

With these results, we can finally calculate $\zeta$,
\begin{eqnarray}
\zeta&=& N - \hat{N} \notag \\
     &=& \hat{N}(\bar{\rho},\varphi_*^a(\xx))-\hat{N}(\bar{\rho}, \bar{\varphi}_*^a) \notag \\
     &=&\hat{N}_a \delta \varphi^a_{\ast}(\xx)
        + \frac{1}{2}\hat{N}_{ab} \delta \varphi^a_{\ast}(\xx) \delta \varphi^b_{\ast}(\xx)
        + \dots   ,
\label{eq:deltaN}
\end{eqnarray}
where $\delta\varphi_*^a(\xx)\equiv \varphi_*^a(\xx)-\bar{\varphi}_*^a$ denotes perturbations to initial scalar fields computed in the flat gauge, and $\hat{N}_a$ and $\hat{N}_{ab}$ are defined as
\begin{equation}
		\hat{N}_a \equiv \frac{\partial \hat{N}[\bar{\rho},\bar{\varphi}_{\ast}^b]}
{\partial \bar{\varphi}_{\ast}^a} , \ \ \ 
\hat{N}_{ab} \equiv \frac{\partial^2 \hat{N}[\bar{\rho},\bar{\varphi}_{\ast}^c]}
{\partial \bar{\varphi}_{\ast}^a \partial \bar{\varphi}_{\ast}^b}  .
		\label{}
\end{equation}
This is the $\delta N$ formalism, which enables us to relate $\zeta$ to
the initial scalar-field perturbations (i.e., scalar-field perturbations
at the initial time) computed in the flat gauge, once we know
derivatives of the number of $e$-folds with respect to the initial
values of the unperturbed  scalar fields $\varphi_*^I$ and their derivatives $\varphi_{N*}^I$.

\subsection{Momentum constraint} 
\label{sec:momentum}
Perhaps a striking thing about the $\delta N$ formalism is that we only had to use the Hamiltonian constraint [Eq.~(\ref{0th-hami})] and the scalar-field equation [Eq.~(\ref{0th-motion-2})] in the gradient expansion. But, should not we also impose the momentum constraint for consistent calculations?

As the momentum constraint comes with a spatial derivative $\partial_i$ we need to consider the momentum constraint in $\mathcal{O}(\epsilon^2)$ in order to derive the correct relationship between physical quantities up to $\mathcal{O}(\epsilon)$,
\begin{equation}
		\partial_i \tilde{H} = -\frac{\tilde{H}}{2M_p^2} G_{IJ}\varphi_N^I \partial_i \varphi^J  + {\cal O}(\epsilon^3) ,
		\label{moment}
\end{equation}
where we have used the fact that $\tilde{A}_{ij} = {\cal O}(\epsilon^2)$.

On the other hand, the Hamiltonian constraint [Eq.~(\ref{0th-hami})] may be differentiated by $\partial_i$ to give 
\begin{equation}
		\partial_i \tilde{H} = -\frac{\tilde{H}}{2M_p^2} G_{IJ}\varphi_N^I \partial_i \varphi^J + B_i  ,
		 \label{B-moment}
\end{equation}
where
\begin{equation}
		B_i \equiv \frac{\tilde{H}^3}{2V}G_{IJ}\left( \varphi^I_N \partial_i \varphi^J_N - \varphi_{NN}^I\partial_i \varphi^J \right) .
		\label{definition_of_B}
\end{equation}
Here, we have used the equation of motion for scalar fields given by Eq.~(\ref{0th-motion-2}), as well as the evolution equation for $K$ given by Eq.~(\ref{full-K-evolution}), which yields $\tilde{H}_N = - \frac{\tilde{H}}{2M_p^2} G_{IJ} \varphi_N^I \varphi_N^J$ in the gradient expansion in the flat gauge.  

Comparing Eqs.~(\ref{B-moment}) and (\ref{moment}), we find that the
$\delta N$ formalism, which does not use the momentum constraint but
uses only the Hamiltonian constraint, can introduce an error
in the momentum constraint by an amount $B_i$. Imposing the
momentum constraint gives an additional constraint $B_i= {\cal O}(\epsilon^3)$ for the $\delta N$ formalism.

How important is $B_i$? In order to investigate the behavior of $B_i$, let us take a spatial derivative of the equation of motion for scalar fields
\begin{align}
		&\tilde{H}\partial_N(\tilde{H} \partial_i \varphi_N^I) +  3\tilde{H}^2 \partial_i \varphi^I_N \notag \\
		&+ \Bigg[ V_{IJ} 
		- \frac{\tilde{H}}{a^3 M_p^2} \frac{d}{dN}\left(G_{JK} a^3 \tilde{H} \varphi^I_N \varphi^K_N \right) \Bigg] 
		\partial_i \varphi^J \notag \\
		& \hspace{2cm} +    H\varphi^I (B_{iN} + 3B_i )   + 2 \varphi_{NN}^I H B_i = 0  .
		\label{aa2}
\end{align}
By contracting this equation with $\varphi^I_N$, one finds
\begin{align}
		& \partial_N B_i +3B_i = 0  ,\notag \\ 
		& \to B_i = \frac{a_{\ast}^3 B_{i \ast}}{a^3}.
		\label{B}
\end{align}
Therefore, $B_i$ has only a decaying solution.\footnote{In linear theory, this quantity is equal to
$-a^{-3}\partial_i W$ where $W$ is given by Eq.~(2.26) of
\cite{Sasaki:1998}, as well as to $-\partial_i \dot{f}$ where $\dot{f}$
is given by Eq.~(5.23) of \cite{Kodama/Hamazaki:1998}. They show
$W\propto a^{-3}$ and $\dot{f}\propto a^{-3}$ in linear theory.}
This is  good news for the $\delta N$ formalism: while it does not {\it
a priori} respect the momentum constraint, the error in the
momentum constraint rapidly decays away by inflation.  
That $B_i$ is a decaying mode may be traced back to the fact that the traceless part of the extrinsic curvature $\tilde{A}_{ij}$ is a decaying mode in the gradient expansion in the flat gauge. In other words, it is a consequence of the universe behaving like a Friedmann-Lema\^itre-Robertson-Walker universe on superhorizon scales, which is guaranteed by Eq.~(\ref{estimation}).

Remember that ignoring the decaying-mode terms of $\beta^i$ and $h_{ij}^{(T)}$ has led to the $\delta N$ formalism and the separate universe description. 
This means that the decaying term $B_i$ should also be ignored for
consistency and thus should be treated as higher order in
$\epsilon$. In fact, the momentum constraint naturally satisfies this 
condition: $B_i = {\cal O}(\epsilon^3)$. 
However, as the $\delta N$ formalism does not {\it a priori} respect the
momentum constraint, it yields the correct growing solutions and
incorrect decaying solutions. In other words, the $\delta N$ formalism
yields valid solutions only in models in which the decaying-mode terms
never affect  the curvature perturbation.
If one needs to completely remove the decaying-mode contributions from
the $\delta N$ formalism, then one should use the $\delta N$ formalism with the initial
condition $B_{i*}=0$. 

One may wonder why consistency between the momentum constraint and
the Hamiltonian constraint gives a relation only among scalar fields
$G_{IJ}\left( \varphi^I_N \partial_i \varphi^J_N -
\varphi_{NN}^I\partial_i \varphi^J \right)=0$, rather than a relation
between the metric variables and scalar fields. This is because we have
ignored the decaying solutions of $\beta^i$ and $\dot{h}^{(T)}_{ij}$. To
see this, let us work at the first order in perturbations, and bring
$\beta^i$ back into Einstein's field equations. We find that consistency between the momentum constraint and the Hamiltonian constraint gives
\begin{equation}
		B_i + \frac{M_p^2 H^2}{V} \partial_i(\partial_j \beta^j)  = 0.
		\label{linear_B}
\end{equation}
Indeed, consistency gives a relation between scalar fields
(contained in $B_i$) and a metric variable ($\beta^i$)\footnote{We
thank M. Sasaki for clarifying this point.}. This equation
also indicates that when we keep the decaying quantities of order
$a^{-3}$, the gradient-expansion order of $\beta^i$ is indeed ${\cal
O}(\epsilon^{-1})$ [see Eq.~(\ref{first-estimation})], as
Eq.~(\ref{linear_B}) gives $\partial_j \beta^j={\cal O}(1)$.

Now is the time to answer the following question: what if we demand
$\beta^i = {\cal O}(\epsilon^0)$? In this case,
Eq.~(\ref{linear_B}) gives $B_i={\cal O}(\epsilon^2)$ or
\begin{equation}
		G_{IJ} \left(  \bar{\varphi}^I_N \delta \varphi^J_N -
		\bar{\varphi}_{NN}^I \delta \varphi^J \right) = {\cal O}(\epsilon),
		\label{}
\end{equation}
where $\delta \varphi^I \equiv \varphi^I - \bar{\varphi}^I$ is the
perturbation of scalar fields in linear theory.
This result indicates that we
are not allowed to have a configuration of scalar fields which yields $G_{IJ} \left(  \bar{\varphi}^I_N \delta \varphi^J_N -
		\bar{\varphi}_{NN}^I \delta \varphi^J \right)={\cal
		O}(\epsilon^0)$.
On the other hand, if we start with
		$\beta^i = {\cal O}(\epsilon^{-1})$, then we can show
		that $G_{IJ} \left(  \bar{\varphi}^I_N \delta \varphi^J_N -
		\bar{\varphi}_{NN}^I \delta \varphi^J \right)$ becomes
		negligible as it is a decaying mode. Another way of
		saying this is that if we start with
		$\beta^i = {\cal O}(\epsilon^0)$, then $G_{IJ} \left(  \bar{\varphi}^I_N \delta \varphi^J_N -
		\bar{\varphi}_{NN}^I \delta \varphi^J \right)=0$ gives
		only one solution for $\delta\varphi^I$, and another
		solution, which corresponds to a decaying mode, does not
		exist.\footnote{Assuming
		$\beta^i={\cal O(\epsilon)}$, Kodama and
		Hamazaki \cite{Kodama/Hamazaki:1998} also find this
		property from the momentum constraint. See their
		Eq.~(5.13) and the argument given below it.}
 		Therefore, demanding that the number of
		independent solutions (which is two) not reduce, one
		should start with $\beta^i = {\cal 
		O}(\epsilon^{-1})$.

\subsection{Slow-roll conditions and momentum constraint}
Interestingly, we can show that $B_i$ vanishes when the slow-roll conditions are satisfied. The slow-roll equations of motion for the canonical
scalar fields are 
\begin{equation}
		3M_p^2 \tilde{H}^2 \approx V , \ \ \ 3\tilde{H}^2 \varphi_N^I + V_I \approx 0 .
		\label{}
\end{equation}
This implies that 
\begin{align}
	  \varphi_N^I \approx - M_p^2 \frac{V_I}{V}  
	 , \ \ \ \ \   \varphi_{NN}^I \approx (2 \varepsilon_{IJ}  - \eta_{IJ})\varphi_N^J ,
		\label{slow-roll}
\end{align}
where the slow-roll parameters $\varepsilon_{ab}$ and $\eta_{ab}$ are defined as  
\begin{equation}
		\varepsilon_{IJ} \equiv \frac{M_p^2}{2}\frac{V_IV_J}{V^2} ,\ \ 
		\eta_{IJ} \equiv M_p^2 \frac{V_{IJ}}{V}.
		\label{}
\end{equation}

Then we find that the following relation is satisfied under the slow-roll condition,
\begin{align}
		 \varphi^I_N \partial_i \varphi^I_N
		\approx \varphi_N^I (2\varepsilon_{IJ} - \eta_{IJ})
 \partial_i \varphi^J \approx \varphi_{NN}^I \partial_i \varphi^I,
\end{align}
which yields $B_i \approx 0$. In this sense, the slow-roll conditions are equivalent to the momentum constraint.  

What does this imply? This implies that the $\delta N$ formalism happens to respect the momentum constraint if the slow-roll conditions are satisfied at the initial time $t_*$. This may provide a partial explanation as to why the $\delta N$ formalism has been successful in computing $\zeta$ for a wide variety of slow-roll inflation models.

\section{Conclusion}
The necessary and sufficient condition for the validity of
the $\delta N$ formalism is that, with a suitable choice of coordinates, the perturbed Hamiltonian constraint and matter-field equations on large scales coincide with the corresponding unperturbed equations.

That perturbed solutions in the long-wavelength limit
can be obtained from unperturbed solutions was found and
investigated by pioneering work in 1998
\cite{Taruya/Nambu:1998,*Nambu/Taruya:1998,Kodama/Hamazaki:1998,Sasaki:1998}. While
their work was restricted to linear theory (and to quasilinear theory
\cite{Sasaki:1998}), we have extended their work to include nonlinear
(but still 
perturbative) perturbations. Such extension is also explored by
\cite{David/etal:2005}, 
who use the so-called ``separate universe approach'' 
\cite{Wands/etal:2000}. As we have
described in Sec.~\ref{sec:previous}, our starting point is more general
than theirs.

In this paper, using the flat gauge $(\psi= C_i = 0)$ and choosing the
number of  $e$-folds $N$ as our time coordinates, we have shown that
the perturbed Hamiltonian constraint and matter-field equations on large scales coincide with the corresponding unperturbed equations,
as long as (at least) the following
conditions are satisfied: 
\begin{itemize}
\item[1.] The unperturbed metric is given by
			   a homogeneous
			  and isotropic
			  Friedmann-Lema\^itre-Robertson-Walker metric.
\item[2.] The final results for the curvature perturbation are not
	  affected by decaying-mode terms, such as the shift vector, a
	  time derivative of tensor perturbations to $g_{ij}/a^2(t)$, or
	  the error in the momentum constraint $B_i$.
 \item[3.] Evolution of scalar-field perturbations outside the horizon
	   can be treated using the lowest order or the next order of
	   the gradient expansion.  
\end{itemize}
In order to show that the shift vector and a time derivative of tensor
perturbations are decaying modes, one needs two more conditions.
\begin{itemize}
\item[4.] Matter fields are given by scalar fields with an arbitrary
	  form of Lagrangian [whose anisotropic stress is 
	  of order ${\cal O}(\epsilon^2)$] or,   
	  in the
	  $n$th-order gradient expansion where $n=0$ or 1, by some
	  fluids with anisotropic stress of order ${\cal
	  O}(\epsilon^{n+1})$.
\item[5.] Theory of gravitation determining the physics of inflation is
	  Einstein's theory, or modified gravitational theories with can
	  be transformed into Einstein's theory.
\end{itemize}
The fourth and fifth conditions are sufficient conditions: we had to use Einstein's field equations to explicitly show that the fourth
condition implies 
that the shift vector and a time derivative
of tensor perturbations are decaying modes. It is possible that other
theories of gravitation require different conditions for the shift vector
and a time derivative of tensor perturbations to be decaying modes.

The discussion in this paper should also apply to vector-field models
(see \cite{Dimastrogiovanni/etal:2010} for a review and references
therein), as long as their anisotropic stress is of order 
${\cal O}(\epsilon^{n+1})$ in the $n$th-order gradient
expansion where $n=0$ or 1.

The third condition is naturally expected in any inflation
scenario. However, when the slow-roll conditions  are violated, it is
known that decaying-mode solutions in the second order of the gradient
expansion cannot be neglected in the power spectrum
\cite{Samuel.M.Leach/M.Sasaki/etal:2001}. In such cases, since there are
no gauges in which a Friedmann-Lema\^itre-Robertson-Walker universe can
be obtained in the second order of the gradient expansion, we can no
longer use the $\delta N$ formalism
\cite{Takamizu/etal:2010,*Takamizu:2011}. 

When all of the above conditions are satisfied, one can calculate $\zeta$ using initial scalar-field perturbations computed in the flat gauge and derivatives of the number of $e$-folds with respect to the initial values of the unperturbed scalar fields $\varphi_*^I$ and their derivatives $\varphi_{N*}^I$.

\begin{acknowledgments}
We would like to thank Y. Itoh, D. Lyth, J. Meyers, A. Naruko, T. Tanaka,
 M. Sasaki, and S. Weinberg for useful discussions and comments. 
We would also like to thank T. Hamazaki for bringing
 Refs.~\cite{Kodama/Hamazaki:1998,Hamazaki:2008} to our attention after
 we have posted 
 our article to arXiv.
This work is supported in part by the GCOE Program ``Weaving Science Web beyond Particle-Matter Hierarchy'' at Tohoku University, a Grant-in-Aid for Scientific Research from JSPS (Grants. No. 18072001 and No. 20540245 for T. F.), the Core-to-Core Program ``International Research Network for Dark Energy,'' as well as by NSF Grant No. PHY-0758153. T. F. would like to thank Luc Branchet and the Institute of Astronomical Observatory, Paris, and N. S. S. and E. K. would like to thank M. Sasaki and the Yukawa Institute for Theoretical Physics, for their warm hospitality during the last stage of this work. 
\end{acknowledgments}

\appendix
\section{Einstein's field equations}
In this Appendix, we give Einstein's field equations in terms of the variables of the ADM formalism [see Eq.~(1) for the ADM metric]. We use 
latin indices for the 3D spatial components running from 1 to 3, and greek indices for the 4D spacetime components running from 0 to 3.

We decompose the three-space metric tensor as 
\begin{equation}
		\gamma_{ij} \equiv a^2(t)e^{2\psi}\tilde{\gamma}_{ij} ,
		\label{}
\end{equation}
where $a(t)$ is the scale factor.  We define $\tilde{\gamma}_{ij}$ such that  $\det[\tilde{\gamma}_{ij}] = 1$; thus,  $\det[\tilde{\gamma}_{ij}]$ can be written as  $\det[(e^{h})_{ij}] = e^{ {\rm Tr}[h]} $ with  ${\rm Tr}[h] = 0$.  We further decompose $h_{ij}$ as 
\begin{equation}
		h_{ij} = \partial_i C_j + \partial_j C_i - \frac{2}{3} \delta_{ij} \partial_k C^k + h^{(T)}_{ij},
		\label{h-deco}
\end{equation}
where $h^{(T)}_{ij}$ denotes a tensor mode and $C_i$ has a scalar mode and a vector mode.

The extrinsic curvature $K_{ij}$ is defined as
\begin{equation}
		K_{ij} \equiv - \nabla_i n_j  = \frac{1}{2\alpha}\left( D_i \beta_j+ D_j \beta_i - \dot{ \gamma}_{ij} \right)  ,
   \label{}
\end{equation}
where $n^{\mu} = \left( 1/\alpha , - \beta^i/\alpha\right)$
is the unit vector normal to the $t$-constant hypersurface, and
$\nabla$ and $D$ are the covariant differential operators constructed by
using $g_{ \mu \nu}$ and $\gamma_{ij}$, respectively. The dots denote time derivatives with respect to $t$.

It is useful to decompose the extrinsic curvature $K_{ij}$ into 
a ``trace'' part $K$ and a ``trace free'' part $\tilde{A}_{ij}$ as
\begin{equation}
K_{ij} =\frac{\gamma_{ij}}{3}K + a^2e^{2\psi}\tilde{A}_{ij} ,
\label{}
\end{equation}
where the indices of $\tilde{A}_{ij}$ are raised/lowered by $\tilde{\gamma}_{ij}$, and $\tilde{A}^i_{\ i} = 0$ is satisfied.

By using the above notations, we write down Einstein's field equations. The Hamiltonian constraint is
\begin{equation}
		R^{(3)} - \tilde{A}_{ij}\tilde{A}^{ij} + \frac{2}{3} K^2 = \frac{2}{M_p^2}T_{\mu\nu}n^{\nu}n^{\mu} .
	\label{full_hami}
\end{equation}
The momentum constraint is
\begin{equation}
		D_j\tilde{A}^j_{\ i} - \frac{2}{3}\partial_i K = - \frac{1}{M_p^2} T_{i\nu} n^{\nu}  .
	\label{full-moment}
\end{equation}
The dynamical equation for $\psi$ is
\begin{equation}
       (\partial_t - \beta^i\partial_i) \psi + H = \frac{1}{3}(- \alpha K + \partial_i \beta^i)  .
       \label{full-K}
\end{equation}
The dynamical equation for $\tilde{\gamma}_{ij}$ is
\begin{equation}
        (\partial_t - \beta^k\partial_k)\tilde{\gamma}_{ij} = -2\alpha\tilde{A}_{ij} + \tilde{\gamma}_{ik}\partial_j \beta^k 
        + \tilde{\gamma}_{jk}\partial_i \beta^k - \frac{2}{3}\tilde{\gamma}_{ij} \partial_k \beta^k ,
	\label{full-gamma}
\end{equation}
which yields the dynamical equation for $K$,
\begin{align}
       (\partial_t - \beta^k\partial_k)K = &\alpha\left( \tilde{A}_{ij}\tilde{A}^{ij} + \frac{1}{3}K^2 \right)  \notag \\
	 & - \gamma^{ij}D_iD_j \alpha + \frac{\alpha}{2M_p^2} \left( T_{\mu\nu}n^{\mu}n^{\nu} + \gamma^{ij}T_{ij} \right),
       \label{full-K-evolution}
\end{align}
as well as the dynamical equation for $\tilde{A}_{ij}$,
\begin{align}
       & (\partial_t - \beta^k\partial_k)\tilde{A}_{ij}  \notag \\
	    = & \frac{1}{a^2e^{2\psi}}\bigg[ \alpha \bigg( R^{(3)}_{ij} - \frac{\gamma_{ij}}{3}R^{(3)}\bigg)
          - \bigg( D_iD_j \alpha - \frac{\gamma_{ij}}{3}D_kD^k \alpha \bigg) \bigg]   \notag\\
         & +  \alpha(K \tilde{A}_{ij} - 2\tilde{A}_{ik}\tilde{A}^k_{\ j})   
		  + \tilde{A}_{ik}\partial_j\beta^k + \tilde{A}_{jk}\partial_i \beta^k - \frac{2}{3}\tilde{A}_{ij}\partial_k \beta^k   \notag\\
		  & \hspace{3cm} - \frac{\alpha}{a^2 e^{2\psi} M_p^2}\bigg( T_{ij} - \frac{\gamma_{ij}}{3}\gamma^{kl} T_{kl} \bigg)  .
	\label{full-A-evolusion}
\end{align}
Here, $T_{ij} - \frac{\gamma_{ij}}{3}\gamma^{kl} T_{kl}$ is the anisotropic stress. 

The three-dimensional Ricci scalar $R^{(3)}$ is constructed from $\gamma_{ij}$, and $M_p^2$ is the reduced Plank mass defined as $1/8\pi G$ where $G$ is the gravitational constant.

\bibliographystyle{apsrev4-1}

\begin{thebibliography}{26}%
\makeatletter
\providecommand \@ifxundefined [1]{%
 \@ifx{#1\undefined}
}%
\providecommand \@ifnum [1]{%
 \ifnum #1\expandafter \@firstoftwo
 \else \expandafter \@secondoftwo
 \fi
}%
\providecommand \@ifx [1]{%
 \ifx #1\expandafter \@firstoftwo
 \else \expandafter \@secondoftwo
 \fi
}%
\providecommand \natexlab [1]{#1}%
\providecommand \enquote  [1]{``#1''}%
\providecommand \bibnamefont  [1]{#1}%
\providecommand \bibfnamefont [1]{#1}%
\providecommand \citenamefont [1]{#1}%
\providecommand \href@noop [0]{\@secondoftwo}%
\providecommand \href [0]{\begingroup \@sanitize@url \@href}%
\providecommand \@href[1]{\@@startlink{#1}\@@href}%
\providecommand \@@href[1]{\endgroup#1\@@endlink}%
\providecommand \@sanitize@url [0]{\catcode `\\12\catcode `\$12\catcode
  `\&12\catcode `\#12\catcode `\^12\catcode `\_12\catcode `\%12\relax}%
\providecommand \@@startlink[1]{}%
\providecommand \@@endlink[0]{}%
\providecommand \url  [0]{\begingroup\@sanitize@url \@url }%
\providecommand \@url [1]{\endgroup\@href {#1}{\urlprefix }}%
\providecommand \urlprefix  [0]{URL }%
\providecommand \Eprint [0]{\href }%
\providecommand \doibase [0]{http://dx.doi.org/}%
\providecommand \selectlanguage [0]{\@gobble}%
\providecommand \bibinfo  [0]{\@secondoftwo}%
\providecommand \bibfield  [0]{\@secondoftwo}%
\providecommand \translation [1]{[#1]}%
\providecommand \BibitemOpen [0]{}%
\providecommand \bibitemStop [0]{}%
\providecommand \bibitemNoStop [0]{.\EOS\space}%
\providecommand \EOS [0]{\spacefactor3000\relax}%
\providecommand \BibitemShut  [1]{\csname bibitem#1\endcsname}%
\let\auto@bib@innerbib\@empty
\bibitem [{\citenamefont {Komatsu}(2010)}]{Komatsu:2010}%
  \BibitemOpen
  \bibfield  {author} {\bibinfo {author} {\bibfnamefont {E.}~\bibnamefont
  {Komatsu}},\ }\href {\doibase 10.1088/0264-9381/27/12/124010} {\bibfield
  {journal} {\bibinfo  {journal} {Class. Quant. Grav.}\ }\textbf {\bibinfo
  {volume} {27}},\ \bibinfo {pages} {124010} (\bibinfo {year} {2010})},\
  \Eprint {http://arxiv.org/abs/1003.6097} {arXiv:1003.6097} \BibitemShut
  {NoStop}%
\bibitem [{\citenamefont {{Desjacques}}\ and\ \citenamefont
  {{Seljak}}(2010)}]{Desjacques/Seljak:2010}%
  \BibitemOpen
  \bibfield  {author} {\bibinfo {author} {\bibfnamefont {V.}~\bibnamefont
  {{Desjacques}}}\ and\ \bibinfo {author} {\bibfnamefont {U.}~\bibnamefont
  {{Seljak}}},\ }\href {\doibase 10.1088/0264-9381/27/12/124011} {\bibfield
  {journal} {\bibinfo  {journal} {Classical and Quantum Gravity}\ }\textbf
  {\bibinfo {volume} {27}},\ \bibinfo {pages} {124011} (\bibinfo {year}
  {2010})},\ \Eprint {http://arxiv.org/abs/1003.5020} {arXiv:1003.5020}
  \BibitemShut {NoStop}%
\bibitem [{\citenamefont {Starobinsky}(1982)}]{Starobinsky:1982}%
  \BibitemOpen
  \bibfield  {author} {\bibinfo {author} {\bibfnamefont {A.~A.}\ \bibnamefont
  {Starobinsky}},\ }\href {\doibase 10.1016/0370-2693(82)90541-X} {\bibfield
  {journal} {\bibinfo  {journal} {Phys. Lett.}\ }\textbf {\bibinfo {volume}
  {B117}},\ \bibinfo {pages} {175} (\bibinfo {year} {1982})}\BibitemShut
  {NoStop}%
\bibitem [{\citenamefont {Starobinsky}(1985)}]{Starobinsky:1986}%
  \BibitemOpen
  \bibfield  {author} {\bibinfo {author} {\bibfnamefont {A.~A.}\ \bibnamefont
  {Starobinsky}},\ }\href@noop {} {\bibfield  {journal} {\bibinfo  {journal}
  {JETP Lett.}\ }\textbf {\bibinfo {volume} {42}},\ \bibinfo {pages} {152}
  (\bibinfo {year} {1985})}\BibitemShut {NoStop}%
\bibitem [{\citenamefont {Salopek}\ and\ \citenamefont
  {Bond}(1990)}]{Salopek/Bond:1990}%
  \BibitemOpen
  \bibfield  {author} {\bibinfo {author} {\bibfnamefont {D.~S.}\ \bibnamefont
  {Salopek}}\ and\ \bibinfo {author} {\bibfnamefont {J.~R.}\ \bibnamefont
  {Bond}},\ }\href {\doibase 10.1103/PhysRevD.42.3936} {\bibfield  {journal}
  {\bibinfo  {journal} {Phys. Rev.}\ }\textbf {\bibinfo {volume} {D42}},\
  \bibinfo {pages} {3936} (\bibinfo {year} {1990})}\BibitemShut {NoStop}%
\bibitem [{\citenamefont {Sasaki}\ and\ \citenamefont
  {Stewart}(1996)}]{Sasaki/Stewart:1996}%
  \BibitemOpen
  \bibfield  {author} {\bibinfo {author} {\bibfnamefont {M.}~\bibnamefont
  {Sasaki}}\ and\ \bibinfo {author} {\bibfnamefont {E.~D.}\ \bibnamefont
  {Stewart}},\ }\href {\doibase 10.1143/PTP.95.71} {\bibfield  {journal}
  {\bibinfo  {journal} {Prog. Theor. Phys.}\ }\textbf {\bibinfo {volume}
  {95}},\ \bibinfo {pages} {71} (\bibinfo {year} {1996})},\ \Eprint
  {http://arxiv.org/abs/astro-ph/9507001} {arXiv:astro-ph/9507001} \BibitemShut
  {NoStop}%
\bibitem [{\citenamefont {Lyth}\ \emph {et~al.}(2005)\citenamefont {Lyth},
  \citenamefont {Malik},\ and\ \citenamefont {Sasaki}}]{David/etal:2005}%
  \BibitemOpen
  \bibfield  {author} {\bibinfo {author} {\bibfnamefont {D.~H.}\ \bibnamefont
  {Lyth}}, \bibinfo {author} {\bibfnamefont {K.~A.}\ \bibnamefont {Malik}}, \
  and\ \bibinfo {author} {\bibfnamefont {M.}~\bibnamefont {Sasaki}},\ }\href
  {\doibase 10.1088/1475-7516/2005/05/004} {\bibfield  {journal} {\bibinfo
  {journal} {JCAP}\ }\textbf {\bibinfo {volume} {0505}},\ \bibinfo {pages}
  {004} (\bibinfo {year} {2005})},\ \Eprint
  {http://arxiv.org/abs/astro-ph/0411220v3} {arXiv:astro-ph/0411220v3}
  \BibitemShut {NoStop}%
\bibitem [{\citenamefont {Lyth}\ and\ \citenamefont
  {Rodriguez}(2005)}]{Lyth/Rodriguez:2005}%
  \BibitemOpen
  \bibfield  {author} {\bibinfo {author} {\bibfnamefont {D.~H.}\ \bibnamefont
  {Lyth}}\ and\ \bibinfo {author} {\bibfnamefont {Y.}~\bibnamefont
  {Rodriguez}},\ }\href {\doibase 10.1103/PhysRevLett.95.121302} {\bibfield
  {journal} {\bibinfo  {journal} {Phys.Rev.Lett.}\ }\textbf {\bibinfo {volume}
  {95}},\ \bibinfo {pages} {121302} (\bibinfo {year} {2005})},\ \Eprint
  {http://arxiv.org/abs/astro-ph/0504045} {arXiv:astro-ph/0504045 [astro-ph]}
  \BibitemShut {NoStop}%
\bibitem [{\citenamefont {Wands}\ \emph {et~al.}(2000)\citenamefont {Wands},
  \citenamefont {Malik}, \citenamefont {Lyth},\ and\ \citenamefont
  {Liddle}}]{Wands/etal:2000}%
  \BibitemOpen
  \bibfield  {author} {\bibinfo {author} {\bibfnamefont {D.}~\bibnamefont
  {Wands}}, \bibinfo {author} {\bibfnamefont {K.~A.}\ \bibnamefont {Malik}},
  \bibinfo {author} {\bibfnamefont {D.~H.}\ \bibnamefont {Lyth}}, \ and\
  \bibinfo {author} {\bibfnamefont {A.~R.}\ \bibnamefont {Liddle}},\ }\href
  {\doibase 10.1103/PhysRevD.62.043527} {\bibfield  {journal} {\bibinfo
  {journal} {Phys.Rev.}\ }\textbf {\bibinfo {volume} {D62}},\ \bibinfo {pages}
  {043527} (\bibinfo {year} {2000})},\ \Eprint
  {http://arxiv.org/abs/astro-ph/0003278} {arXiv:astro-ph/0003278 [astro-ph]}
  \BibitemShut {NoStop}%
\bibitem [{\citenamefont {Dimopoulos}\ \emph {et~al.}(2009)\citenamefont
  {Dimopoulos}, \citenamefont {Karciauskas}, \citenamefont {Lyth},\ and\
  \citenamefont {Rodriguez}}]{Dimopoulos/etal:2009}%
  \BibitemOpen
  \bibfield  {author} {\bibinfo {author} {\bibfnamefont {K.}~\bibnamefont
  {Dimopoulos}}, \bibinfo {author} {\bibfnamefont {M.}~\bibnamefont
  {Karciauskas}}, \bibinfo {author} {\bibfnamefont {D.~H.}\ \bibnamefont
  {Lyth}}, \ and\ \bibinfo {author} {\bibfnamefont {Y.}~\bibnamefont
  {Rodriguez}},\ }\href {\doibase 10.1088/1475-7516/2009/05/013} {\bibfield
  {journal} {\bibinfo  {journal} {JCAP}\ }\textbf {\bibinfo {volume} {0905}},\
  \bibinfo {pages} {013} (\bibinfo {year} {2009})},\ \Eprint
  {http://arxiv.org/abs/0809.1055} {arXiv:0809.1055 [astro-ph]} \BibitemShut
  {NoStop}%
\bibitem [{\citenamefont {Arnowitt}\ \emph {et~al.}(1962)\citenamefont
  {Arnowitt}, \citenamefont {Deser},\ and\ \citenamefont {Misner}}]{ADM:1962}%
  \BibitemOpen
  \bibfield  {author} {\bibinfo {author} {\bibfnamefont {R.~L.}\ \bibnamefont
  {Arnowitt}}, \bibinfo {author} {\bibfnamefont {S.}~\bibnamefont {Deser}}, \
  and\ \bibinfo {author} {\bibfnamefont {C.~W.}\ \bibnamefont {Misner}},\
  }\href@noop {} {\  (\bibinfo {year} {1962})},\ \bibinfo {note} {gravitation:
  an introduction to current research, Louis Witten ed. (Wilew 1962), chapter
  7, pp 227-265},\ \Eprint {http://arxiv.org/abs/gr-qc/0405109}
  {arXiv:gr-qc/0405109 [gr-qc]} \BibitemShut {NoStop}%
\bibitem [{\citenamefont {Maldacena}(2003)}]{Maldacena:2003}%
  \BibitemOpen
  \bibfield  {author} {\bibinfo {author} {\bibfnamefont {J.~M.}\ \bibnamefont
  {Maldacena}},\ }\href@noop {} {\bibfield  {journal} {\bibinfo  {journal}
  {JHEP}\ }\textbf {\bibinfo {volume} {05}},\ \bibinfo {pages} {013} (\bibinfo
  {year} {2003})},\ \Eprint {http://arxiv.org/abs/astro-ph/0210603}
  {astro-ph/0210603} \BibitemShut {NoStop}%
\bibitem [{\citenamefont {Hamazaki}(2008)}]{Hamazaki:2008}%
  \BibitemOpen
  \bibfield  {author} {\bibinfo {author} {\bibfnamefont {T.}~\bibnamefont
  {Hamazaki}},\ }\href {\doibase 10.1103/PhysRevD.78.103513} {\bibfield
  {journal} {\bibinfo  {journal} {Phys.Rev.}\ }\textbf {\bibinfo {volume}
  {D78}},\ \bibinfo {pages} {103513} (\bibinfo {year} {2008})},\ \Eprint
  {http://arxiv.org/abs/0811.2366} {arXiv:0811.2366 [astro-ph]} \BibitemShut
  {NoStop}%
\bibitem [{\citenamefont {Lyth}\ and\ \citenamefont
  {Liddle}(2009)}]{Lyth/Liddle:2009}%
  \BibitemOpen
  \bibfield  {author} {\bibinfo {author} {\bibfnamefont {D.~H.}\ \bibnamefont
  {Lyth}}\ and\ \bibinfo {author} {\bibfnamefont {A.~R.}\ \bibnamefont
  {Liddle}},\ }\href@noop {} {\emph {\bibinfo {title} {{The primordial density
  perturbation: Cosmology, inflation and the origin of structure}}}}\ (\bibinfo
   {publisher} {Cambridge University Press},\ \bibinfo {address} {Oxford, UK},\
  \bibinfo {year} {2009})\BibitemShut {NoStop}%
\bibitem [{\citenamefont {Weinberg}(2008)}]{Weinberg1}%
  \BibitemOpen
  \bibfield  {author} {\bibinfo {author} {\bibfnamefont {S.}~\bibnamefont
  {Weinberg}},\ }\href {\doibase 10.1103/PhysRevD.78.123521} {\bibfield
  {journal} {\bibinfo  {journal} {Phys.Rev.}\ }\textbf {\bibinfo {volume}
  {D78}},\ \bibinfo {pages} {123521} (\bibinfo {year} {2008})},\ \Eprint
  {http://arxiv.org/abs/0808.2909} {arXiv:0808.2909 [hep-th]} \BibitemShut
  {NoStop}%
\bibitem [{\citenamefont {Weinberg}(2009)}]{Weinberg2}%
  \BibitemOpen
  \bibfield  {author} {\bibinfo {author} {\bibfnamefont {S.}~\bibnamefont
  {Weinberg}},\ }\href {\doibase 10.1103/PhysRevD.79.043504} {\bibfield
  {journal} {\bibinfo  {journal} {Phys.Rev.}\ }\textbf {\bibinfo {volume}
  {D79}},\ \bibinfo {pages} {043504} (\bibinfo {year} {2009})},\ \Eprint
  {http://arxiv.org/abs/0810.2831} {arXiv:0810.2831 [hep-ph]} \BibitemShut
  {NoStop}%
\bibitem [{\citenamefont {Sasaki}\ and\ \citenamefont
  {Tanaka}(1998)}]{Sasaki:1998}%
  \BibitemOpen
  \bibfield  {author} {\bibinfo {author} {\bibfnamefont {M.}~\bibnamefont
  {Sasaki}}\ and\ \bibinfo {author} {\bibfnamefont {T.}~\bibnamefont
  {Tanaka}},\ }\href {\doibase 10.1143/PTP.99.763} {\bibfield  {journal}
  {\bibinfo  {journal} {Prog.Theor.Phys.}\ }\textbf {\bibinfo {volume} {99}},\
  \bibinfo {pages} {763} (\bibinfo {year} {1998})},\ \Eprint
  {http://arxiv.org/abs/gr-qc/9801017} {arXiv:gr-qc/9801017} \BibitemShut
  {NoStop}%
\bibitem [{\citenamefont {Bardeen}\ \emph {et~al.}(1983)\citenamefont
  {Bardeen}, \citenamefont {Steinhardt},\ and\ \citenamefont
  {Turner}}]{Bardeen/Turner/Steinhardt:1983}%
  \BibitemOpen
  \bibfield  {author} {\bibinfo {author} {\bibfnamefont {J.~M.}\ \bibnamefont
  {Bardeen}}, \bibinfo {author} {\bibfnamefont {P.~J.}\ \bibnamefont
  {Steinhardt}}, \ and\ \bibinfo {author} {\bibfnamefont {M.~S.}\ \bibnamefont
  {Turner}},\ }\href {\doibase 10.1103/PhysRevD.28.679} {\bibfield  {journal}
  {\bibinfo  {journal} {Phys. Rev.}\ }\textbf {\bibinfo {volume} {D28}},\
  \bibinfo {pages} {679} (\bibinfo {year} {1983})}\BibitemShut {NoStop}%
\bibitem [{\citenamefont {Weinberg}(2003)}]{Weinberg:2003}%
  \BibitemOpen
  \bibfield  {author} {\bibinfo {author} {\bibfnamefont {S.}~\bibnamefont
  {Weinberg}},\ }\href {\doibase 10.1103/PhysRevD.67.123504} {\bibfield
  {journal} {\bibinfo  {journal} {Phys.Rev.}\ }\textbf {\bibinfo {volume}
  {D67}},\ \bibinfo {pages} {123504} (\bibinfo {year} {2003})},\ \Eprint
  {http://arxiv.org/abs/astro-ph/0302326} {arXiv:astro-ph/0302326 [astro-ph]}
  \BibitemShut {NoStop}%
\bibitem [{\citenamefont {Kodama}\ and\ \citenamefont
  {Hamazaki}(1998)}]{Kodama/Hamazaki:1998}%
  \BibitemOpen
  \bibfield  {author} {\bibinfo {author} {\bibfnamefont {H.}~\bibnamefont
  {Kodama}}\ and\ \bibinfo {author} {\bibfnamefont {T.}~\bibnamefont
  {Hamazaki}},\ }\href {\doibase 10.1103/PhysRevD.57.7177} {\bibfield
  {journal} {\bibinfo  {journal} {Phys.Rev.}\ }\textbf {\bibinfo {volume}
  {D57}},\ \bibinfo {pages} {7177} (\bibinfo {year} {1998})},\ \Eprint
  {http://arxiv.org/abs/gr-qc/9712045} {arXiv:gr-qc/9712045 [gr-qc]}
  \BibitemShut {NoStop}%
\bibitem [{\citenamefont {Taruya}\ and\ \citenamefont
  {Nambu}(1998)}]{Taruya/Nambu:1998}%
  \BibitemOpen
  \bibfield  {author} {\bibinfo {author} {\bibfnamefont {A.}~\bibnamefont
  {Taruya}}\ and\ \bibinfo {author} {\bibfnamefont {Y.}~\bibnamefont {Nambu}},\
  }\href {\doibase 10.1016/S0370-2693(98)00378-5} {\bibfield  {journal}
  {\bibinfo  {journal} {Phys.Lett.}\ }\textbf {\bibinfo {volume} {B428}},\
  \bibinfo {pages} {37} (\bibinfo {year} {1998})},\ \Eprint
  {http://arxiv.org/abs/gr-qc/9709035} {arXiv:gr-qc/9709035 [gr-qc]}
  \BibitemShut {NoStop}%
\bibitem [{\citenamefont {Nambu}\ and\ \citenamefont
  {Taruya}(1998)}]{Nambu/Taruya:1998}%
  \BibitemOpen
  \bibfield  {author} {\bibinfo {author} {\bibfnamefont {Y.}~\bibnamefont
  {Nambu}}\ and\ \bibinfo {author} {\bibfnamefont {A.}~\bibnamefont {Taruya}},\
  }\href {\doibase 10.1088/0264-9381/15/9/021} {\bibfield  {journal} {\bibinfo
  {journal} {Class.Quant.Grav.}\ }\textbf {\bibinfo {volume} {15}},\ \bibinfo
  {pages} {2761} (\bibinfo {year} {1998})},\ \Eprint
  {http://arxiv.org/abs/gr-qc/9801021} {arXiv:gr-qc/9801021 [gr-qc]}
  \BibitemShut {NoStop}%
\bibitem [{\citenamefont {Dimastrogiovanni}\ \emph {et~al.}(2010)\citenamefont
  {Dimastrogiovanni}, \citenamefont {Bartolo}, \citenamefont {Matarrese},\ and\
  \citenamefont {Riotto}}]{Dimastrogiovanni/etal:2010}%
  \BibitemOpen
  \bibfield  {author} {\bibinfo {author} {\bibfnamefont {E.}~\bibnamefont
  {Dimastrogiovanni}}, \bibinfo {author} {\bibfnamefont {N.}~\bibnamefont
  {Bartolo}}, \bibinfo {author} {\bibfnamefont {S.}~\bibnamefont {Matarrese}},
  \ and\ \bibinfo {author} {\bibfnamefont {A.}~\bibnamefont {Riotto}},\ }\href
  {\doibase 10.1155/2010/752670} {\bibfield  {journal} {\bibinfo  {journal}
  {Adv.Astron.}\ }\textbf {\bibinfo {volume} {2010}},\ \bibinfo {pages}
  {752670} (\bibinfo {year} {2010})},\ \Eprint {http://arxiv.org/abs/1001.4049}
  {arXiv:1001.4049 [astro-ph.CO]} \BibitemShut {NoStop}%
\bibitem [{\citenamefont {Leach}\ \emph {et~al.}(2001)\citenamefont {Leach},
  \citenamefont {Sasaki}, \citenamefont {Wands},\ and\ \citenamefont
  {Liddle}}]{Samuel.M.Leach/M.Sasaki/etal:2001}%
  \BibitemOpen
  \bibfield  {author} {\bibinfo {author} {\bibfnamefont {S.~M.}\ \bibnamefont
  {Leach}}, \bibinfo {author} {\bibfnamefont {M.}~\bibnamefont {Sasaki}},
  \bibinfo {author} {\bibfnamefont {D.}~\bibnamefont {Wands}}, \ and\ \bibinfo
  {author} {\bibfnamefont {A.~R.}\ \bibnamefont {Liddle}},\ }\href {\doibase
  10.1103/PhysRevD.64.023512} {\bibfield  {journal} {\bibinfo  {journal}
  {Phys.Rev.}\ }\textbf {\bibinfo {volume} {D64}},\ \bibinfo {pages} {023512}
  (\bibinfo {year} {2001})},\ \Eprint {http://arxiv.org/abs/astro-ph/0101406}
  {arXiv:astro-ph/0101406} \BibitemShut {NoStop}%
\bibitem [{\citenamefont {Takamizu}\ \emph {et~al.}(2010)\citenamefont
  {Takamizu}, \citenamefont {Mukohyama}, \citenamefont {Sasaki},\ and\
  \citenamefont {Tanaka}}]{Takamizu/etal:2010}%
  \BibitemOpen
  \bibfield  {author} {\bibinfo {author} {\bibfnamefont {Y.}~\bibnamefont
  {Takamizu}}, \bibinfo {author} {\bibfnamefont {S.}~\bibnamefont {Mukohyama}},
  \bibinfo {author} {\bibfnamefont {M.}~\bibnamefont {Sasaki}}, \ and\ \bibinfo
  {author} {\bibfnamefont {Y.}~\bibnamefont {Tanaka}},\ }\href {\doibase
  10.1088/1475-7516/2010/06/019} {\bibfield  {journal} {\bibinfo  {journal}
  {JCAP}\ }\textbf {\bibinfo {volume} {1006}},\ \bibinfo {pages} {019}
  (\bibinfo {year} {2010})},\ \Eprint {http://arxiv.org/abs/1004.1870}
  {arXiv:1004.1870} \BibitemShut {NoStop}%
\bibitem [{\citenamefont {Takamizu}\ and\ \citenamefont
  {Yokoyama}(2011)}]{Takamizu:2011}%
  \BibitemOpen
  \bibfield  {author} {\bibinfo {author} {\bibfnamefont {Y.}~\bibnamefont
  {Takamizu}}\ and\ \bibinfo {author} {\bibfnamefont {J.}~\bibnamefont
  {Yokoyama}},\ }\href {\doibase 10.1103/PhysRevD.83.043504} {\bibfield
  {journal} {\bibinfo  {journal} {Phys.Rev.}\ }\textbf {\bibinfo {volume}
  {D83}},\ \bibinfo {pages} {043504} (\bibinfo {year} {2011})},\ \Eprint
  {http://arxiv.org/abs/1011.4566} {arXiv:1011.4566} \BibitemShut {NoStop}%
\end{thebibliography}
%
\end{document}